\documentclass[prx,twocolumn,showpacs,superscriptaddress]{revtex4-1}
\usepackage[english]{babel}
\usepackage[dvips]{graphicx}
\usepackage{color}
\usepackage{latexsym}
\usepackage{amsmath}
\usepackage{amsthm}
\usepackage{amsfonts}
\usepackage{amssymb}
\usepackage{mathtools}
\usepackage{array}
\usepackage{cancel}
\usepackage{algorithmic}
\usepackage[dvipsnames]{xcolor}

\newcolumntype{P}[1]{>{\centering\arraybackslash}p{#1}}

\begin{document}

\pagenumbering{arabic}

\title{Engineering the speedup of quantum tunneling in Josephson systems via dissipation}
\author{D.\ Maile}
\affiliation{Institut f\"ur komplexe Quantensysteme,  Universit{\"a}t Ulm, D-89069 Ulm, Germany}

\author{J. Ankerhold}
\affiliation{Institut f\"ur komplexe Quantensysteme,  Universit{\"a}t Ulm, D-89069 Ulm, Germany}
\author{S. Andergassen}
\affiliation{Institut f\"ur Theoretische Physik and Center for Quantum Science, Universit{\"a}t T{\"u}bingen, Auf der Morgenstelle 14, 72076 T{\"u}bingen, Germany}
\author{W. Belzig}
\affiliation{Fachbereich Physik, Universit{\"a}t Konstanz, D-78457 Konstanz,  Germany}

\author{G. Rastelli}
\affiliation{INO-CNR BEC Center and Dipartimento di Fisica, Universit{\`a} di Trento, I-38123 Povo, Italy}
\begin{abstract}
We theoretically investigate the escape rate occurring via quantum tunneling in a system affected by tailored dissipation.
Specifically, we study the environmental assisted quantum tunneling
of the superconducting phase  in a current-biased Josephson junction.
We consider Ohmic resistors inducing dissipation both in the phase and in the charge of the quantum circuit. 
We find that the charge dissipation leads to an enhancement of the quantum escape rate.
This effect appears already in the low Ohmic regime and also occurs in the presence of phase dissipation that favors localization.
Inserting realistic circuit parameters, we address the question of its experimental observability and discuss suitable parameter spaces for the observation of the enhanced rate. 
\end{abstract}
\date{\today}
\maketitle

%
%
%
%
\section{Introduction}
The inevitable coupling of a quantum system to its surrounding environment is a major obstacle for the accessibility and usability of quantum behavior for quantum technologies.
Over the last decades, however, the perception of dissipative couplings between the bath and the system changed, and the openness of quantum systems has first been proposed \cite{mirrahimi_dynamically_2014,braun_creation_2002,gilles_generation_1994,poyatos_quantum_1996}, and later been shown \cite{dassonneville_dissipative_2021,leghtas_confining_2015,shankar_autonomously_2013} to be also a 
resource. 
In particular, by engineering coherent driving in presence of specific reservoirs, open systems can yield steady target states exhibiting protected quantum behavior \cite{zapletal_stabilization_2022}. 
%
%
In these systems the environmental degrees of freedom are a resource for quantum behavior, however, the mentioned effects strongly rely on the coherent control of the system and the specific form of the environment.

Here, we propose a system that exploits the environment as a resource by using basic Ohmic environments, without any further coherent control.
Specifically, we engineer a quantum system that uses the environment to speedup 
the escape rate occurring via quantum tunneling, an effect that is at the heart of many quantum technologies, especially in superconducting circuits \cite{krantz_quantum_2019,wendin_quantum_2017}.
%
%
%
%
%
%
%
\begin{figure}[b]
	\centering
	\includegraphics[scale=0.57]{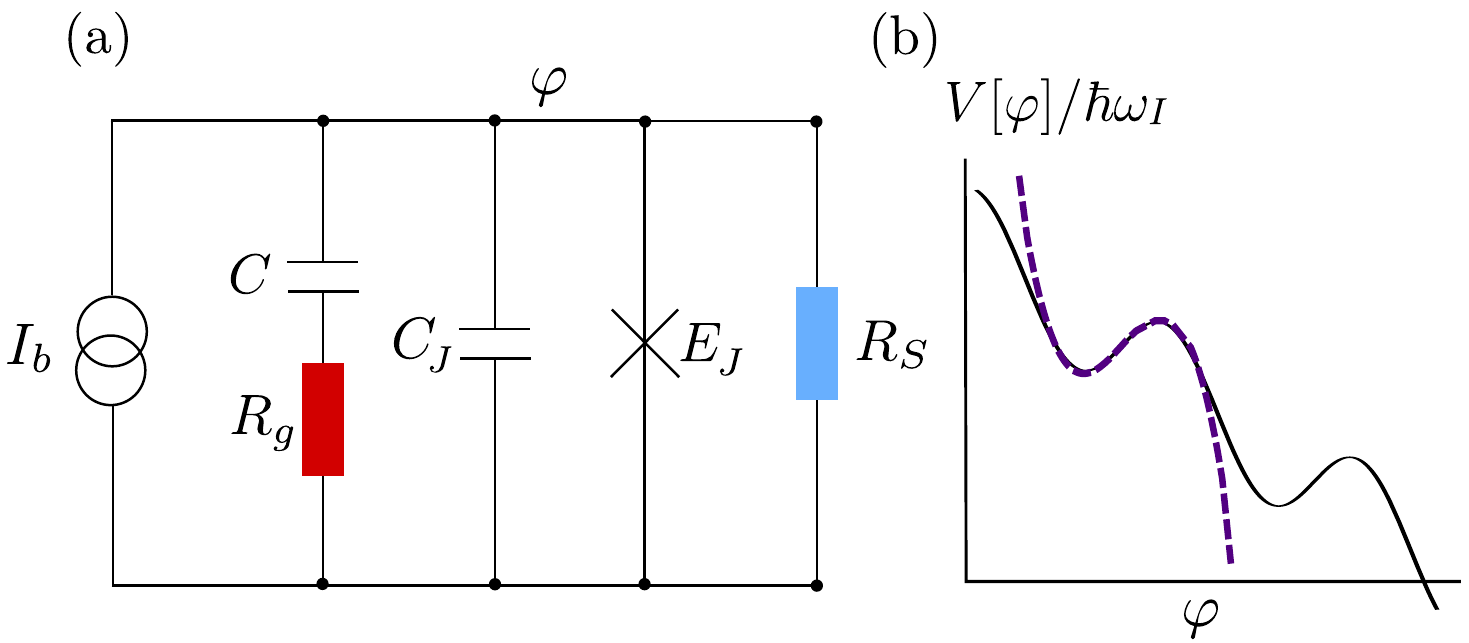}
	\caption{ (a) Scheme of the current-biased Josephson junction. 
	The capacitance $C_J$ is the intrinsic capacitance of the Josephson junction 
	which is connected in parallel to a shunt resistor $R_S$ and 
	and to a branch containing an external capacitance $C$ and a resistance $R_g$.
	(b) Sketch of the tilted cosine potential (solid line) $V[\varphi]$ for the superconducting phase difference across the junction and its cubic approximation (dashed line).
	}
	\label{Fig.1}
\end{figure}
%
%
%
The first measurements of macroscopic quantum tunneling of the superconducting phase  in a current-biased Josephson junction \cite{devoret_measurements_1985}, paved the way for nowadays superconducting qubits \cite{martinis_rabi_2002,yu_coherent_2002,martinis_quantum_2020}.
Already in their seminal article \cite{devoret_measurements_1985}, Devoret et al. included dissipation in their theoretical formula for the tunneling rate. 
They modeled the dissipation via a resistor shunting the junction that leads to an exponential suppression of tunneling. 
This result directly follows from the Caldeira-Leggett model, where the detrimental effect of the dissipation on the tunneling process was calculated via path integral methods \cite{caldeira_quantum_1983,caldeira_quantum_1984}. 
However, the dissipative coupling favors localization only because it couples to the  phase of the junction. 
This coupling can be seen as a weak measurement of the phase reducing its uncertainty and therefore the tunneling rate. 
Since the phase and the charge of a Josephson junction are conjugate variables in the quantum regime, at zero temperature the uncertainty relation must be fulfilled.
Then, if one manages to couple the charge to a dissipative bath, the charge fluctuations are suppressed with the concomitant enhancement of the phase fluctuations.
Ultimately, this can lead to a larger tunneling rate when, for instance, the phase evolves in an effective energy potential with several minima.

The described theoretical framework was already studied in the more abstract context of dissipative position and momentum couplings 
\cite{leggett_quantum_1984,cuccoli_quantum_2001,kohler_dissipative_2006,ankerhold_dissipation_2007,cuccoli_reentrant_2010,maile_effects_2020,maile_exponential_2021,Maile2021_thesis,Kuzyakin2011,Ferialdi2017,rastelli_dissipation-induced_2016}. 
In this work, we engineer the reservoirs for a current-biased Josephson system 
with two distinct dissipative interactions affecting respectively the phase and the charge variable.
In this Josephson circuit, the dynamics of the superconducting phase difference 
is set by the so-called tilted washboard potential having (quantum) metastable minima.
We show that the escape rate of the phase from a metastable minimum 
is enhanced as a function of the dissipative coupling strength to the charge. 
Specifically, we consider the circuit in Fig.~\ref{Fig.1}a, 
where the shunted resistance $R_S$ yields the standard phase dissipative interaction 
and 
the resistance $R_g$ in series with an (external) capacitance $C$ yields a charge dissipative interaction.

We structure our study as follows. 
In Sec.~\ref{Sec.2}, we first introduce the Hamiltonian of the system 
and the theoretical framework that we use to calculate the escape rate.
In Sec.~\ref{Sec.3}, we discuss the nature of the dissipative environments and 
introduce a variational approach to approximately calculate the escape rate 
in presence of dissipation. 
We benchmark the variational method by comparing it with an exact numerical method (more details on the theoretical description of the Ohmic environments and the comparison between the variational and the numerical results are provided in the Appendices). 
Afterwards, we discuss our results on the impact of the dissipative couplings
and work out their dependence 
on the parameters of the Josephson circuit. 
We comment on the experimental observability of our results in Sec.~\ref{Sec.4} and conclude in Sec.\ref{Sec.5}. 
%
%
%
%
\section{Theoretical model and methods}  
\label{Sec.2}
\subsection{Model Hamiltonian of the Josephson junction}
For a current-biased Josephson junction  with critical current $ I_C $, a capacitance $ C_J $, and shunted by a capacitance $ C $ as displayed in Fig.~\ref{Fig.1}a (with $R_g=0$
 and $R_S\rightarrow \infty $), 
the junction Hamiltonian in the quantum regime reads \cite{Schon:1990kj,zagoskin_quantum_2011}
\begin{align}
\label{eq:H_S}
\hat{\mathcal{H}}_{S}
=\frac{\hat{Q}^2}{2C_{tot}}
-
\frac{\hbar I_C}{2e}
\cos\left(\hat{\varphi}\right)-\frac{\Phi_{0} I_{b}}{2\pi }\hat{\varphi},
\end{align}
with $C_{tot}=C_J + C$.
The phase operator 
$\hat{\varphi}$ describes the phase difference between the two superconductors forming the junction and
$\hat{Q}$
is its conjugate charge operator $ [\hat{\varphi},\hat{Q}]=i 2e $, i.e. the charge tunneling 
through the junction. 
$ e $ is the electron charge and
$ \Phi_0=h/(2e) $ the flux quantum.
The second and the third terms in Eq.~(\ref{eq:H_S}) form the tilted washboard potential for the phase $V\left[ \varphi\right]$, see 
Fig.~\ref{Fig.1}b.

Assuming to initialize the system to a state in which the phase is localized around a minimum of the potential, we calculate the decay rate of the phase in the metastable potential.
We focus the analysis in the semiclassical limit, where the height of the potential barrier is $ V_0 \gg \hbar \omega_I $, with $ \omega_I $ being the harmonic frequency of the well \cite{ankerhold-springer2007}. 
In this regime, the ratio between Josephson- and charging energy is of the order of $E_J/E_C\sim 10^5$, with $E_C=4e^2/ C_{tot}$ and $E_{J}=\Phi_0 I_C/(2\pi)$.
In this limit, at zero temperature ($\beta=\hbar/(k_B T)\rightarrow \infty$), the  escape rate takes the form 
\cite{coleman_fate_1977,langer_theory_1967} 
\begin{align}
\Gamma_0 = K_0 \, e^{-S^{(S)}_{B}/\hbar},
\label{Eq.decay}
\end{align}
where $S^{(S)}_{B}=S_S[\varphi_{B}^{(S)}(\tau)] $ is the Euclidean (imaginary time) action of the system 
calculated on the saddle point path $\varphi_{B}^{(S)}(\tau)$, namely the limit path that minimizes the action and that starts and ends at the minimum in the time range $ \tau \in [-\frac{\beta}{2},\frac{\beta}{2}]$ for $\beta\rightarrow \infty$.
The prefactor $K_0$ is related to the Gaussian fluctuations around this path.
The Euclidean action for the junction reads
\begin{align}
\label{eq:ss}
S_S [\varphi(\tau)] &
\!=\!\int_{-\frac{\beta}{2}}^{\frac{\beta }{2}}\!d\tau
\left[ \frac{C_{tot}\Phi_0^2}{8\pi^2}\dot{\varphi}^2(\tau) +
V\left[ \varphi\left( \tau \right) \right]
\right],
\end{align}
with $ \dot{\varphi}=d\varphi/d\tau $. 
Since we are interested in the decay out of a metastable state, we approximate the tilted cosine potential locally with the effective cubic potential
\begin{align}
V \left[ \varphi \right] \simeq \frac{C_{tot}\Phi_{0}^{2}}{4\pi^ 2}\left(\frac{1}{2}\omega_{I}^{2}\varphi^{2}-\frac{1}{3}\omega_{J}^{2}\varphi^{3}\right),
\end{align}
where $\omega_{I}\!=\!\sqrt{\frac{2\pi I_{C}}{C_{tot}\Phi_{0}}}\left(1-\left(I_{b}/I_{C}\right)^{2}\right)^{\frac{1}{4}}$
and $\omega_{J}\!=\!\sqrt{\frac{\pi I_{b}}{C_{tot}\Phi_{0}}}$.
For $I_{b}\rightarrow I_{C}$ the cubic potential is known to serve as an accurate approximation for all practical purposes if $I_{b}/I_{C}\geq0.98$ which is chosen throughout this work. 
Further, the barrier height normalized 
with respect to the frequency of the well is determined by
\begin{equation}
\frac{V_0}{\hbar\omega_{I}}=\frac{\sqrt{2\pi I_{C}C_{tot}\Phi_{0}^{3}}}{6\pi^2\hbar }\frac{\left(1-(I_b/I_C)^{2}\right)^{\frac{5}{4}}}{(I_b/I_C)^{2}}.
\end{equation}
To find the minimizing path $ \varphi_B^{(S)} $, one inserts the ansatz $ \varphi(\tau) = \varphi_B^{(S)}(\tau)+\delta\varphi(\tau) $ into the action \eqref{eq:ss} and sets the terms proportional to $ \delta\varphi(\tau) $ to zero. 
This yields the following equation in 
frequency space
\begin{align}
\left(\omega^{2}\!+\!\omega_{I}^{2}\right)\! \tilde{\varphi}_B^{(S)}(\omega)\!=\!\frac{\omega_{J}^{2}}{2 \pi}\! \int_{-\infty}^{\infty}\!\!\!\!d\omega' \tilde{\varphi}_B^{(S)}\left(\omega\!+\!\omega^{\prime}\right) \tilde{\varphi}_{B}^{(S)}\left(\omega^{\prime}\right)\!,
\label{Eq.nondissdiff}
\end{align}
where $ \tilde{\varphi}_B^{(S)}(\omega)=\int_{-\infty}^{\infty}d\tau \varphi_{B}^{(S)}(\tau) e^{-i\omega \tau}$. The integral Eq.~(\ref{Eq.nondissdiff}) has the known \textit{bounce} solution 
\begin{align}
\varphi_{B}^{(S)}(\tau) & =\frac{3}{2}\frac{\omega_{I}^{2}}{\omega_{J}^{2}}\frac{1}{\cosh^{2}\left(\frac{\omega_{I}\tau}{2}\right)},
\label{Eq.nondissbounce}
\end{align}
with the result 
$
S_B^{(S)}=\frac{108}{15}\frac{V_0}{\hbar\omega_I}. 
$

\subsection{Action in presence of dissipation}

In presence of dissipation, the functional form of the escape rate 
does not change \cite{caldeira_quantum_1983} and is still given by
\begin{align}
\Gamma = K \, e^{-S_{B}/\hbar}.
\label{Eq.decay2}
\end{align}
This rate now describes an environmental assisted quantum tunneling through the barrier.
Here $S_{B}$ is the Euclidean action including the dissipative effect of the environment and evaluated at its corresponding minimizing path
$S_{B}=S [\varphi_B(\tau)] $.
For the circuit shown in Fig.~\ref{Fig.1}a with the resistors, the Euclidean action reads
\begin{align}
\label{eq:Saction}
S [\varphi(\tau)] &
= 
S_S [\varphi(\tau)]\! +\!\frac{1}{2}  \iint_{-\infty }^{\infty}d\tau d\tau' F^{(\varphi)}(\tau-\tau')\varphi(\tau)\varphi(\tau') \nonumber\\ 
&+\frac{1}{2} \iint_{-\infty }^{\infty}d\tau d\tau' F^{(Q)}(\tau-\tau')\dot{\varphi}(\tau)\dot{\varphi}(\tau') .
\end{align}
The second term is related to the shunt resistance 
with the function $F^{(\varphi)}(\tau)$ 
which in frequency space is given by
\begin{equation}
\tilde{F}^{(\varphi)}(\omega)
=
\frac{\Phi_{0}^{2}}{4\pi^2 R_S}  |\omega| \label{Eq.phasekernel}
\,  .
\end{equation}
This kind of phase (Ohmic) dissipation  was the subject of many studies and its impact has been investigated in several systems \cite{caldeira_quantum_1983,Weiss2012}. 
To quantify it, it is useful to introduce the parameter $ \gamma=1/(R_S C_{tot}) $ which represents the damping coefficient of the particle moving in the effective potential.

On the other side, the branch with the resistance $R_g$ in series with the external capacitance $C$ leads to a qualitatively different kind of dissipation.
The effect of these elements are described by the third term in Eq.~(\ref{eq:Saction})
with the 
function (kernel) $F^{(Q)}(\tau) $ which in frequency space reads
\begin{equation}
\tilde{F}^{(Q)}(\omega)
=
-
\frac{C\Phi_{0}^{2}}{4\pi^2}
\frac{\tau_{p}|\omega|}{1+\tau_{p}|\omega|} \,, \label{Eq.chargekernel}
\end{equation}
with $\tau_p=R_gC$ the relaxation time 
associated to the two elements that act as
dissipative couplings. 
Note that the coupling parameter $ \tau_p $ is proportional to the resistance $ R_g $.
This kind of dissipation, that we denote charge dissipation, can lead 
to a suppression of the charge fluctuations \cite{maile_quantum_2018}.
For more information on the theoretical background we refer to the  Appendix~\ref{Ap:dissipation} as well as to Ref. \cite{maile_quantum_2018}.

The corresponding equation for the minimizing path 
$\varphi_B(\omega)$ in Fourier space is given by
\begin{align}
\Big(\omega^{2}\!+\!\omega_{I}^{2}\label{Eq.diffeq} +\!\Big.&\Big.\frac{4\pi^2F^{(\varphi)}(\omega)}{C_{tot}\Phi_0^2}\!+\!\frac{4\pi^2F^{(Q)}(\omega)}{C_{tot}\Phi_0^2}\omega^2\Big) \tilde{\varphi}_B(\omega)=\nonumber\\&\;\;\;\frac{\omega_{J}^{2}}{2 \pi} \int_{-\infty}^{\infty} \!\!d\omega'\,\tilde{\varphi}_B\left(\omega+\omega^{\prime}\right) \tilde{\varphi}_{B}\left(\omega^{\prime}\right). 
\end{align}
From this equation, we can already get some insight into the effect of charge dissipation on the system. By combining the kinetic and the charge dissipation part of Eq.~(\ref{Eq.diffeq}), we find
\begin{equation}
    C_{tot}\omega^2+\frac{4\pi^2F^{(Q)}(\omega)}{\Phi_0^2}\omega^2 \!=\! \omega^2 \left(C_{tot}\!-\!C\frac{\tau_p|\omega|}{1+\tau_p|\omega|}\right)
\end{equation}
and interpret the effect of charge dissipation as an effective reduction of the capacitance of the circuit 
in the high frequency region $|\omega| \tau_p\gg1$.
However, at finite frequency, the charge dissipation has a dynamical effect beyond a simple renormalization of the capacitance.

In order to calculate the bounce solution of the full dissipative problem, we need to solve Eq.~(\ref{Eq.diffeq}). 
Before we discuss its exact numerical solution, we present a variational approach yielding a good approximation to the exact result.

\subsection{Variational bounce solution in presence of dissipation} 
We here present the variational approach used to find the approximate solution $ \varphi_V(\tau)\approx \varphi_B(\tau) $. Our ansatz for this procedure is a modified version of the non-dissipative bounce path of Eq.~(\ref{Eq.nondissbounce}), a similar procedure was performed in Refs.~\cite{freidkin_decay_1986,caldeira_quantum_1983}.
We introduce variational parameters $ {A} $ and $ {B} $ such that 
\begin{align}
\tilde{\varphi}_{V}(\omega) & =\frac{3}{2}\frac{\omega_{I}^{2}}{\omega_{J}^{2}}\int_{-\infty}^{\infty}d\tau\frac{{A}}{\cosh^{2}\left({B}\frac{\omega_{I}\tau}{2}\right)}e^{-i\omega\tau} ,\label{Eq.varbounce}
\end{align}
defines a set of paths with the proper boundary conditions.
Inserting these paths 
into the full dissipative action, we find the minimal action path by minimizing the latter with respect to the parameters $ {A}  $ and $ {B}  $, i.e. we set $ dS_V/d{A}=0 $ and $ dS_V/d{B} = 0 $, to determine 
the extremal solution $(\bar{A},\bar{B})$.
The variational action evaluated at the minima $\bar{A}$ and $\bar{B}$ then reads
\begin{align}
{S}_V&  \!=\!\frac{27V_{0}\bar{A}^2}{2\omega_{I}}\!\left(\!\frac{4{\bar{B}}}{15}\!+\!\frac{4}{3{\bar{B}}}\!-\!\frac{16{\bar{A}}}{15{\bar{B}}}\!+\!\frac{\gamma\kappa}{\omega_{I}}\!-\!\frac{f(\tau_{p},{\bar{B}})}{2(1\!+\!\frac{C_J}{C})}\!\right),
\label{Eq.Sv}
\end{align}
where $ \kappa =12\xi(3)/\pi^3$, with $\xi(x)$ the zeta - function and $ f(\tau_p,\bar{B}) $ given by
\begin{equation}
   f(\tau_p,\bar{B}) =\frac{16\pi}{\bar{B}^{4}}\int_{0}^{\infty}dx\frac{\tau_{p}\omega_{I}x^{5}}{\left(1+\tau_{p}\omega_{I}x\right)}\frac{1}{\sinh^{2}\left(\frac{\pi}{\bar{B}}x\right)}.
\end{equation}
The parameter  $\bar{B}$ satisfies the following equation
\begin{equation}
{\bar{B}}^{2}\left(\frac{4}{9}\!-\!\frac{g(\tau_{p},{\bar{B}})}{(1\!+\!\frac{C_J}{C})}\right)\!+\!{\bar{B}}\left(\frac{2}{3}\frac{\gamma\kappa}{\omega_{I}}\!+\!\frac{5}{3}\frac{f(\tau_{p},{\bar{B}})}{(1\!+\!\frac{C_J}{C})}\right)\!-\!\frac{4}{9}\!=\!0\, ,
\label{Eq.B}
\end{equation}
where we defined the function
\begin{equation}
   g(\tau_p,\bar{B}) =\frac{16\pi}{\bar{B}^{6}}\int_{0}^{\infty}dx\frac{\tau_{p}\omega_{I}x^{6}}{\left(1+\tau_{p}\omega_{I}x\right)}\frac{\coth\left(x\frac{\pi}{\bar{B}}\right)}{\sinh^{2}\left(x\frac{\pi}{\bar{B}}\right)}.
\end{equation}
The parameter $\bar{A}$ is determined by
\begin{equation}
\bar{A}=\frac{5}{16}\left[{\bar{B}}^{2}\frac{8}{15}+\frac{8}{3}+{\bar{B}}\left(2\frac{\gamma\kappa}{\omega_{I}}-\frac{f\left(\tau_{p},{\bar{B}}\right)}{(1\!+\!\frac{C_J}{C})}\right)\right] .
\label{Eq.A}
\end{equation}
 In Eqs.~(\ref{Eq.Sv}-\ref{Eq.B}), we see that the effect of charge dissipation is suppressed by the factor $(1+C_J/C)$. This is due to the fact, that only the charge at $C$ is affected by the resistance $R_g$, as will discussed in more detail in Sec.~\ref{Sec.4} below.

\subsection{Exact bounce solution in presence of dissipation} 
For the calculation of the exact bounce path $ \varphi_{B}(\tau) $, we use the iterative technique introduced by Chang and Chakravarty \cite{chang_quantum_1984}.
To perform the numerical computation, 
we define the rescaled action $\mathcal{S}=8\pi^2S/(C_{tot}\Phi_{0}^{2}\omega_{I}\varphi_{0}^{2})$ depending only on dimensionless quantities, with  $\varphi_0=\frac{3}{2}\frac{\omega_{I}^{2}}{\omega_{J}^{2}}$.
The corresponding rescaled differential equation for the bounce path reads
\begin{align}
z(x)&=D^{-1}(x)\frac{3}{2}\frac{1}{2\pi}\int_{-\infty}^{\infty}dx'\;z(x'+x)\,z(x'),
\label{Eq.numdif}
\end{align}
where
\begin{align}
\!D(x)\!=\!{\left(\!x^{2}\!+\!z(x)\!+\!\frac{\gamma}{\omega_{I}}|x|\!-\!\frac{x^{2}}{(1\!+\!\frac{C_J}{C})}\frac{\tau_{p}|x|\omega_{I}}{\left(1\!+\!\tau_{p}\omega_{I}|x|\right)}\!\right)},
\end{align}
with $x=\omega/\omega_I$ and $z(x)=\tilde{\varphi}_B(x)/\varphi_0$. We discretize Eq.~(\ref{Eq.numdif}) to calculate the convolution numerically and solve the equation iteratively. Following Ref.~\cite{chang_quantum_1984}, we avoid a dangerous direction in the iterative procedure by substituting the factor in front of the integral via $ \lambda_0 = 3/(4\pi)$ for the first ansatz and rescale the obtained $z_1(x,\lambda_{0})$  to $z_1(x,\lambda_{1})$, where $ \lambda_1 = \lambda_{0} (z_0(0)/z_{1}(0))^2$.
Continuing this procedure iteratively until convergence yields the corresponding numerical solution for the bounce path. 
%
%
%
%
%
%
%
\begin{figure}[b]
	\centering
	\includegraphics[scale=0.27]{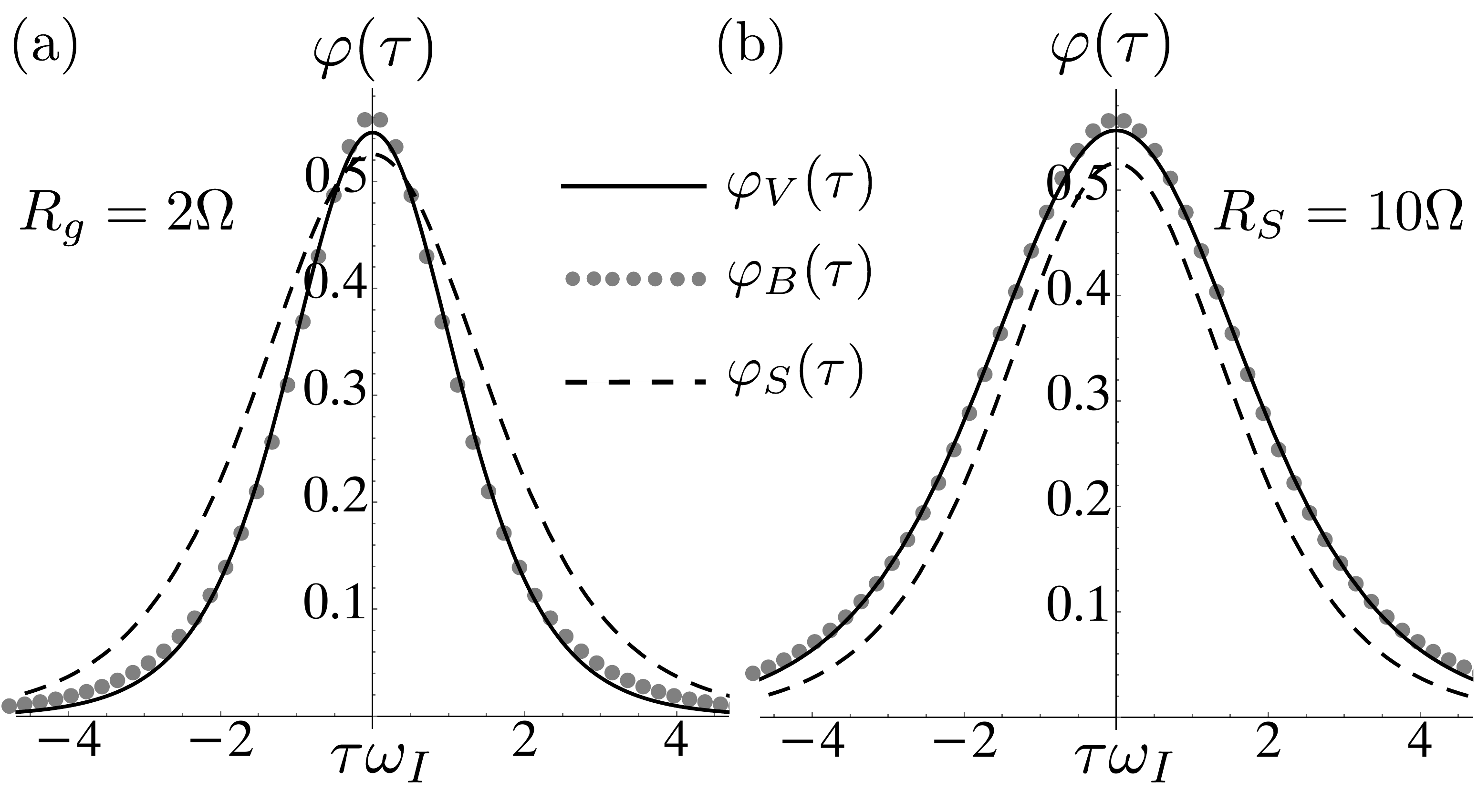}
	\caption{Bounce path in presence of pure charge dissipation $(R_S=\infty)$ in (a) and pure phase dissipation $(R_g=0)$ in (b). $\varphi_V(\tau)$ shows the variational treatment (solid line), $\varphi_B(\tau)$ corresponds to the numerical values (dots), and $\varphi_S(\tau)$ denotes the non-dissipative bounce (dashed line). Parameters: $I_b/I_C = 0.985$, $I_C=21 \, \mu\text{A}$, $C_{tot}=6\, \text{pF}$ and $C_J/C\rightarrow 0$.
	}
	\label{Fig.Fig2b}
\end{figure}

We find that the variational path and 
action reproduce in an excellent way the exact numerical results in the parameters ranges discussed in this work.
As an illustrative example, we report a comparison between the variational bounce and the exact solution in Fig.~\ref{Fig.Fig2b}, both for pure charge and pure phase dissipation, together with the non-dissipative case.
Therefore, hereafter, we consider the variational action for the discussion of 
the results.

%
\section{Results} \label{Sec.3}
We here present results on 
the effects of the resistors on the escape rate. Our analysis is restricted 
to the impact on the exponential part in Eq.~(\ref{Eq.decay2}), which represents the leading contribution.
Specifically, we study the quantity $ \Gamma/\Gamma_0 $, where $ \Gamma_0 $ denotes the decay rate without dissipation, and further approximate the prefactor $ K $ to be independent of the dissipation $ K/K_0 \approx 1$.
This assumption is justified also by 
our findings in a previous work, where we showed that the change in the prefactor does not have a significant influence on the qualitative behavior \cite{maile_exponential_2021}. 

Therefore, we describe the influence of dissipation on the tunneling rate via the quantity 
\begin{align}
\mathcal{E}= e^{-\frac{1}{\hbar}\left(S_V-S_B^{(S)}\right)},
\end{align}
where $ S_B^{(S)}=S_S[\varphi^{(S)}_B] $ is the non-dissipative action on the bounce path and $ S_V $ the variational solution defined in Eqs.~(\ref{Eq.Sv}-\ref{Eq.A}).

For $ \mathcal{E} >1 $, the dissipative couplings speed up the escape through the barrier, while 
they suppress it for $ \mathcal{E} <1 $. 
In the following sections, we analyze the different cases using realistic parameters.
In the presented results, we use a fixed value of $ C_{tot} = 6 \,\text{pF} $ for simplicity. 
However, we will consider changes of the potential form by varying $ I_b $ and $ I_C $ and analyze the limit $C_J/C\rightarrow 0$. 
The influence of the a finite ratio $ C_J/C $ will be discussed in Sec.~\ref{Sec.4}.

\begin{figure*}[t]
	\centering
	\includegraphics[scale=0.345]{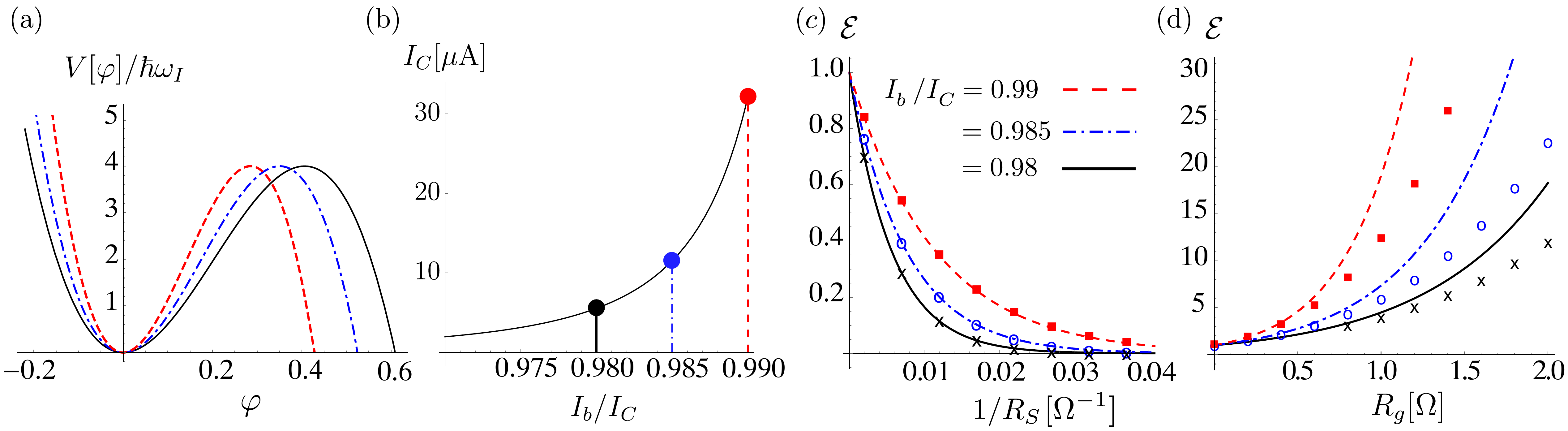}
	\caption{(a) Cubic potential for a fixed ratio $V_0/(\hbar\omega_I)=4$ and different values of $ I_b/I_C $. For larger ratios $I_b/I_C$ the barrier becomes narrower and hence steeper. (b) Values for $I_b$ and $I_C$ for a fixed barrier height. Each color combination of dot and line in (b)
	corresponds to the respective barrier in (a). (c) and (d) Results for $\mathcal{E}$ as a function of dissipation, for fixed $ V_0/(\hbar\omega_I) = 4 $ and different ratios $I_b/I_C$. 
	The respective values for $I_C$ can be read in (b).  
	The symbols display the results using the
	unperturbed path $\varphi^{(S)}_B$ in the dissipative action 
	(undamped-bounce-approximation, see text). 
	Panel (c) is for pure phase dissipation:  suppression of the tunneling as a function of $ R_S^{-1} $.
	And (d) for pure charge dissipation: enhancement of the tunneling as function of $R_g$. Capacitance $C_{tot} = 6\,\text{pF}$ and $C_J/C\rightarrow 0$. 
	}
	\label{Fig.Fig3}
\end{figure*}
%

\subsection{Fixing the barrier height $V_0/(\hbar\omega_I)$}
We first consider a fixed 
barrier height of $ V_0/(\hbar \omega_I) = 4 $, see Fig.~\ref{Fig.Fig3}a for the corresponding potential form.
By 
changing the ratio $ I_b/I_C $, we can adjust the steepness of the potential barrier: larger ratios $I_b/I_C$ lead to narrower (and steeper) barriers.
Note that, in order to keep the barrier height fixed, $I_C$ is different for different ratios $ I_b/I_C $, as shown in Fig.~\ref{Fig.Fig3}b.
%
%
%
%
%
%
%

In Fig.~\ref{Fig.Fig3}, we further show the results for $\mathcal{E}$ for pure phase dissipation in (c) and pure charge dissipation in (d), for the potentials of Fig. \ref{Fig.Fig3}a. 
For pure phase dissipation
$(R_g=0)$ displayed in Fig.~\ref{Fig.Fig3}c, the well-known behavior is reproduced: 
the presence of the shunt resistance $R_S$ yields a suppression of quantum tunneling.
This effect was exposed by Caldeira and Leggett in their seminal work  \cite{caldeira_quantum_1983}.
Apart from using a variational approach to approximate the dissipative bounce path, inserting the unperturbed bounce path $\varphi^{(S)}_B$ into the dissipative action can be used a as a first order correction (see also \cite{caldeira_quantum_1983}).  
%
We see that for pure phase dissipation using the undamped bounce provides a good approximation. This holds also for the variational treatment, since
the exact bounce path is not substantially altered by the presence of the phase dissipation. As a consequence, 
the unperturbed bounce
solution $\varphi^{(S)}_B$ is sufficient to describe the effect of phase dissipation in this regime.
We also find that the effect of phase dissipation becomes more significant for broader barriers, an effect detected also in previous studies \cite{freidkin_decay_1986}.
%
%
%
%
%
%
%
%

For pure charge dissipation, representative 
results are reported in Fig.~\ref{Fig.Fig3}d.
In this case, we find exactly the opposite behavior: 
By increasing the resistance $ R_g $, the escape rate is enhanced.
Moreover, the effect is more pronounced when the barrier becomes steeper.  
Already resistances of a few Ohm ($ R_g \approx 1 \Omega$) lead to a dramatic increase of the escape rate.
%
%
%
%
%
%
\begin{figure}[b]
	\centering
	\includegraphics[scale=0.55]{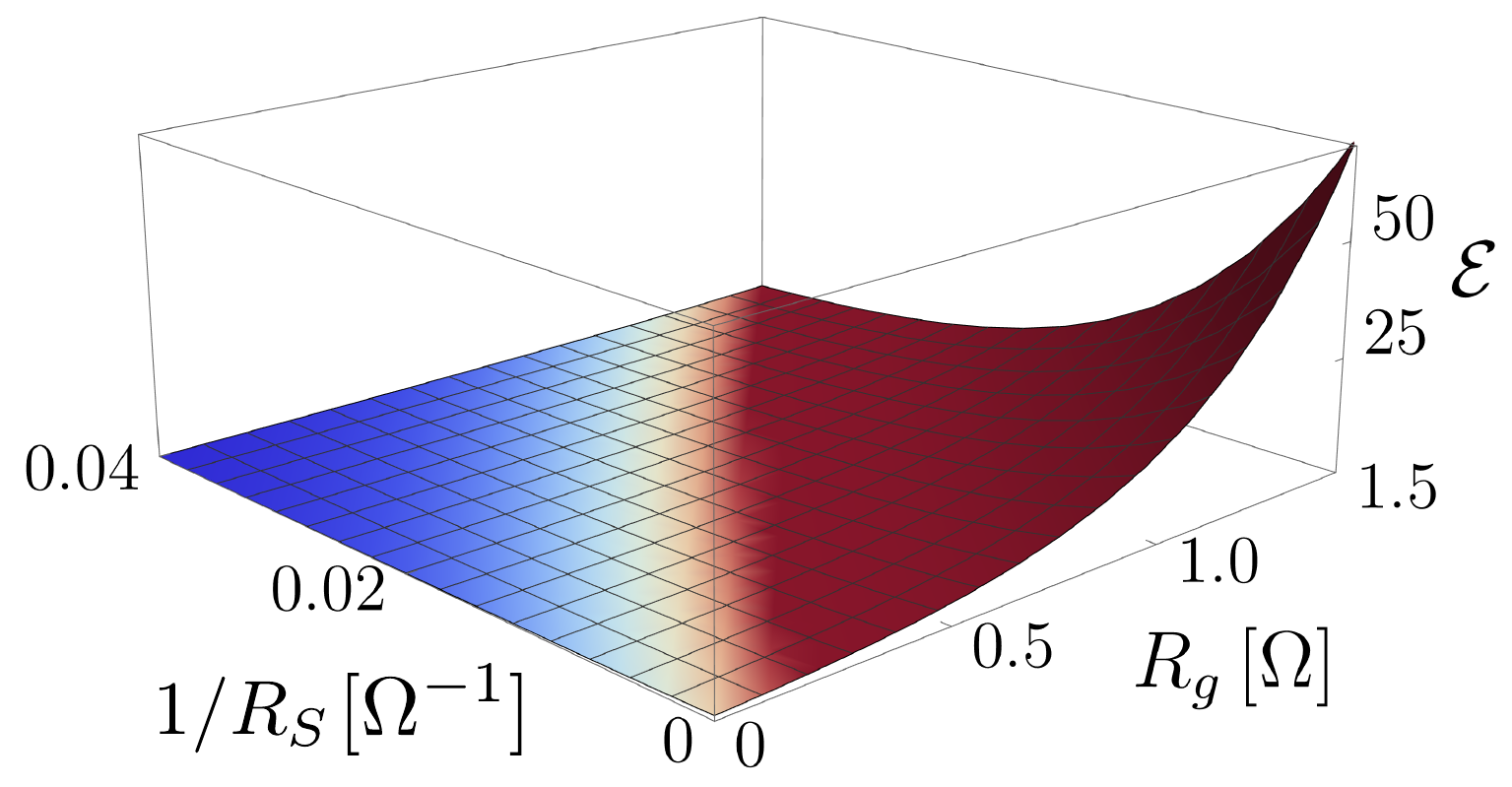}
	\caption{Landscape plot of $\mathcal{E}$ as a function of both resistances $R_S$ and $R_g$ for $ V_0/(\hbar\omega_I) = 4 $. The red area corresponds to the region where the exponential factor is enhanced due to dissipation, the blue are instead where the tunneling rate is suppressed. The other parameters are: $I_b/I_C = 0.99$, $I_C = 32\,\mu$A,  $C_{tot}=6$~pF.
}
	\label{Fig.Fig4}
\end{figure}
We also note that inserting the unperturbed bounce solution $\varphi^{(S)}_B$ in the dissipative action (we call this the undamped-bounce-approximation) results to be a drastic approximation that strongly underestimates the effect of charge dissipation.

In presence of both resistors, we find a large parameter space favoring the speedup of quantum tunneling. 
This 
is illustrated in the landscape plot for $\mathcal{E}$ as a function of both resistances $R_S$ and $R_g$ (for $ V_0/(\hbar\omega_I) = 4 $), see Fig.~\ref{Fig.Fig4}.
The red area corresponds to the region where the exponential factor is enhanced due to dissipation, while the blue one to the region  where the tunneling rate is suppressed.

\subsection{Fixing the critical current $I_C$}
\begin{figure*}[t!]
	\centering
	\includegraphics[scale=0.342]{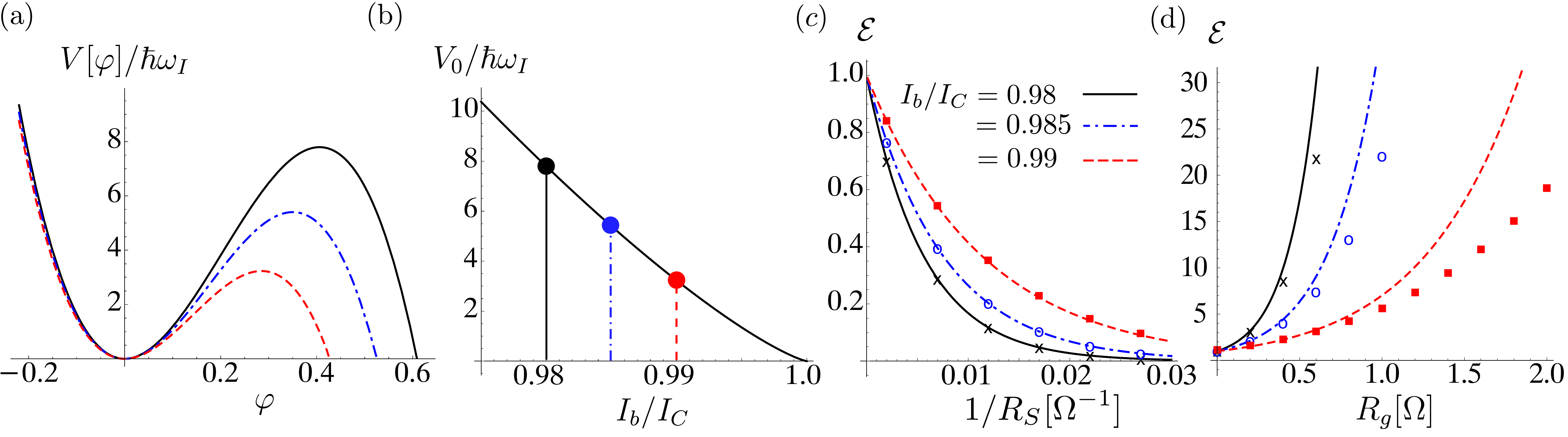}
	\caption{ (a) Change of the barrier for a fixed $ I_C $ and variable $I_b$. (b) Barrier height for different values of $ I_b/I_C $ for fixed $ I_C=21\,\mu\text{A} $. The colors correspond to the barriers displayed in (a).
	(c) and (d) Results for $\mathcal{E}$ for a fixed $ I_C = 21 \mu\text{A}  $ and different ratios $I_b/I_C$.
	The respective values for the ratio $V_0/(\hbar\omega_I)$ can be found in (b). 
   The symbols display the results using the
	unperturbed path $\varphi^{(S)}_B$ in the dissipative action 
	(the undamped-bounce-approximation, see text).
	Panel (c) is for pure phase dissipation: Suppression of the tunneling as a function of $ R_S^{-1}$. 
	And (d) for pure charge dissipation: Enhancement of the tunneling as function of $ R_g $. Capacitance $C_{tot} = 6\,\text{pF}$ and $C_J/C\rightarrow 0$. }
	\label{Fig.Fig6}
\end{figure*}

In order to explore the full parameter space, we also consider different barrier heights.
By varying the ratio $ I_b/I_C $, for a fixed  value of the critical current of $ I_C = 21 $ $\mu\text{A}$, the potential barrier changes as shown in Fig.~\ref{Fig.Fig6}a. We note that all parameters used here are comparable to the experimental values of Ref.~\cite{martinis_rabi_2002}.
%
%
%
%
%
%
%
%
%
%
The corresponding results for $\mathcal{E}$ are reported in Fig.~\ref{Fig.Fig6}, both for pure phase dissipation in (c) and pure charge dissipation in (d). 
For pure phase dissipation, we again find a suppression of the tunneling and that the influence of $R_S$ slightly 
decreases for more shallow 
barriers (red curves). 
As discussed above, 
using 
the unperturbed bounce path $\varphi^{(S)}_B$ in the dissipative action is a good approximation 
in this regime.
%
%
%
%
%
%
%

In Fig.~\ref{Fig.Fig6}d, the results for pure charge dissipation are shown.
Here, we find qualitatively the same behavior for the dissipative influence of $R_g$ which decreases for more shallow barriers. 
However, the enhancement is still significant even for 
$I_b/I_C = 0.99$ that approximately describes the experimental setup and parameters 
of the Josephson circuit used in Ref.~\cite{martinis_rabi_2002}.
We also see that the use of the unperturbed bounce solution in the dissipative action (undamped-bounce-approximation) 
underestimates the effect of charge dissipation, as already noted in the previous section.

%
%
%
%
\begin{figure}[b!]
	\centering
	\includegraphics[scale=0.244]{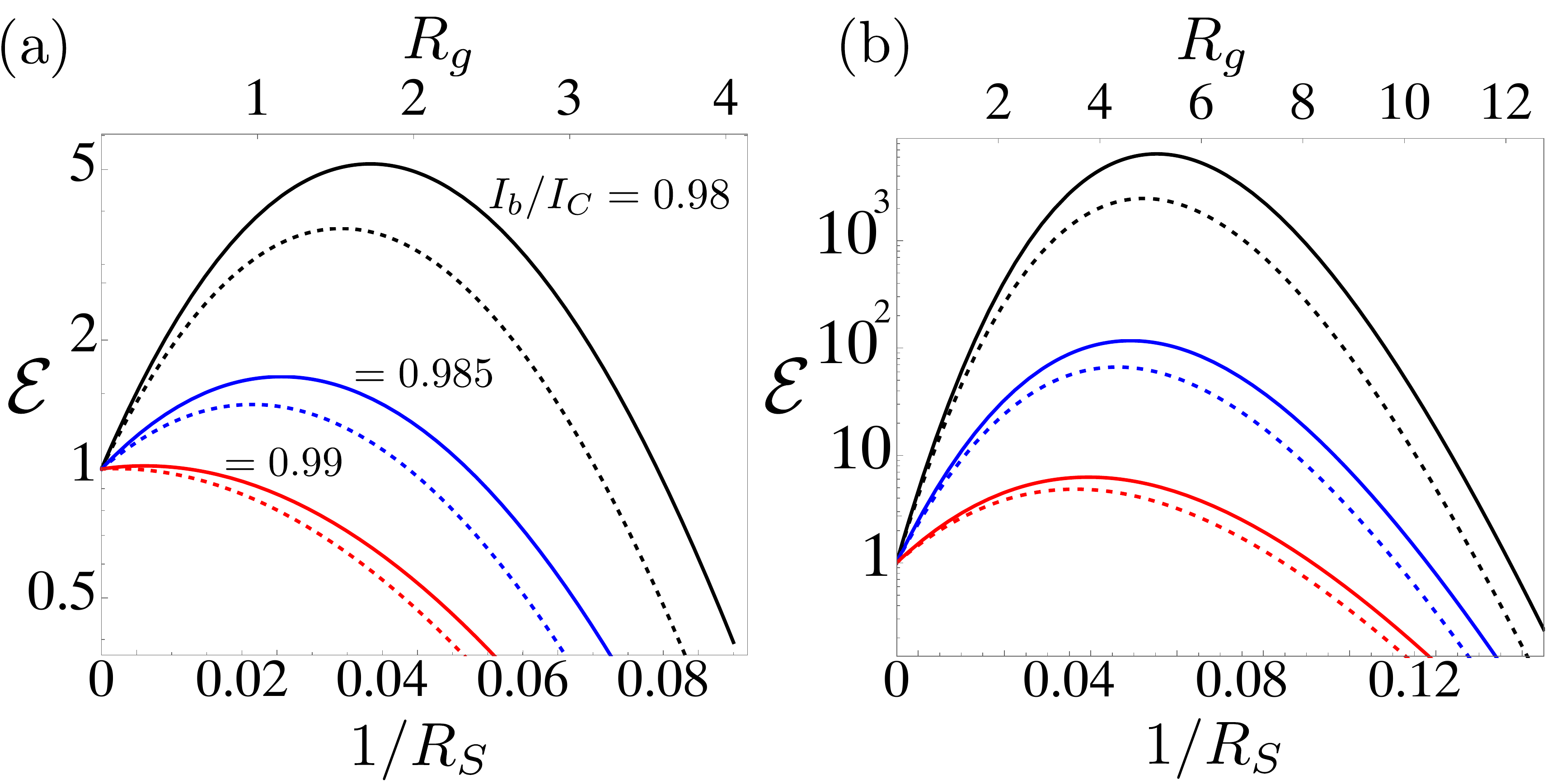}
	\caption{Enhancement of $\mathcal{E}$ 
	in presence of both resistors, for  $\tau_p\omega_I^2/\gamma=\text{const}$.
	The two panels display two different regimes of enhancement (or suppression) of the tunneling:
    from top to bottom $\tau_p\omega_I^2/\gamma=3.43, 2.98, 2.43$ in (a) and $\tau_p\omega_I^2/\gamma=6.47, 5.61, 4.59$ in (b), for $C_J=0$ (solid lines) and $C_J/C=0.02$ (dashed lines).
	The other parameters are: $I_C=21\,\mu$A and $C_{tot}=6$~pF. 
	In (b) for every $1/R_S$, the value $R_g$ is larger than in (a). Therefore, the enhancement is larger in (b) then in (a).  This is also reflected by larger values of $\tau_p\omega_I^2/\gamma$ for the same $I_b/I_C$.
	}
	\label{Fig.boxplots}
\end{figure}

Finally, we discuss the results for having both kinds of dissipation shown in Fig.~\ref{Fig.boxplots} (for  $\tau_p\omega_I^2/\gamma=\text{const}.$). 
Instead of plotting 
$\mathcal{E}$ as a function of the two resistors,  
we illustrate the results by fixing the ratio between the dissipative coupling strenghts in order to identify parameter regimes where  $\mathcal{E}>1$ or $\mathcal{E}<1$. 
We also illustrate the change of $\mathcal{E}$ by varying the values of $I_b/I_C$, and
%
%
compare also the limit 
$C_J=0$ (solid lines) with a finite Josephson capacitance of $C_J/C=0.02$ (dotted lines).
The two panels display two different regimes:
In (b) for every $1/R_S$, the value $R_g$ is larger than in (a). Therefore, the enhancement is larger in (b) then in (a), which is also reflected by larger values of $\tau_p\omega_I^2/\gamma$ for the same $I_b/I_C$.
For the dependence on $C_J$, as long as $C_J\ll C$, the results for a finite Josephson capacitance are comparable to the case $C_J=0$.


%
%
\section{Remarks concerning the experimental realizability} 
\label{Sec.4}
In this section, we critically discuss 
the possibility to experimentally observe the enhancement of the quantum escape rate in Josephson circuits and provide 
some more quantitative information on 
experimental relevant parameters.

\begin{figure}[t]
	\centering
	\includegraphics[scale=0.52]{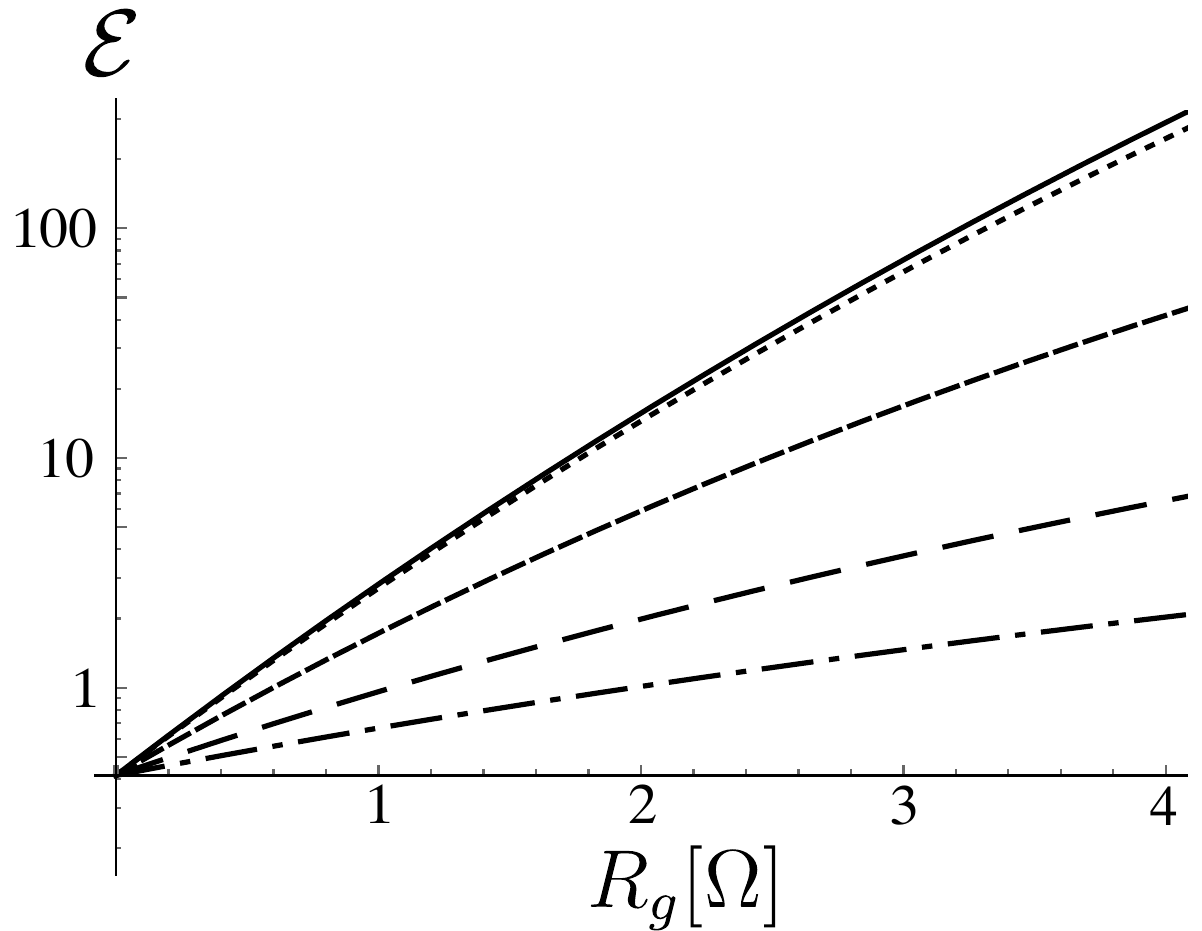}
	\caption{Enhancement of $\mathcal{E}$ as a function of $ R_g $, for different ratios of the two capacitances $ C_J/C =0,0.01,0.2,0.5,1$ from top to bottom.  The other parameters are: $ I_b/I_C = 0.99,\,R_S = 100 \Omega $, $I_C=21\,\mu$A, and $C_{tot}=6$~pF . 
}
	\label{Fig.C-CJ}
\end{figure}

First, we examine the effect of the junction capacitance.
Due to its presence the dissipative interaction only affects a part of the total charge of the circuit and, therefore, the influence of charge dissipation is suppressed.
This is evident in the Fig.~\ref{Fig.C-CJ} where we plot 
$\mathcal{E}$ for different ratios $C_J/C$.
We emphasize that, in a current-biased junction, the regime discussed here  $C \gg C_J$ is a common situation.
%
Even if the two capacitances are comparable $C_J \sim 0.5 C$,
the enhancement of the escape is still significant.

\begin{figure}[t]
	\centering
	\includegraphics[scale=0.5]{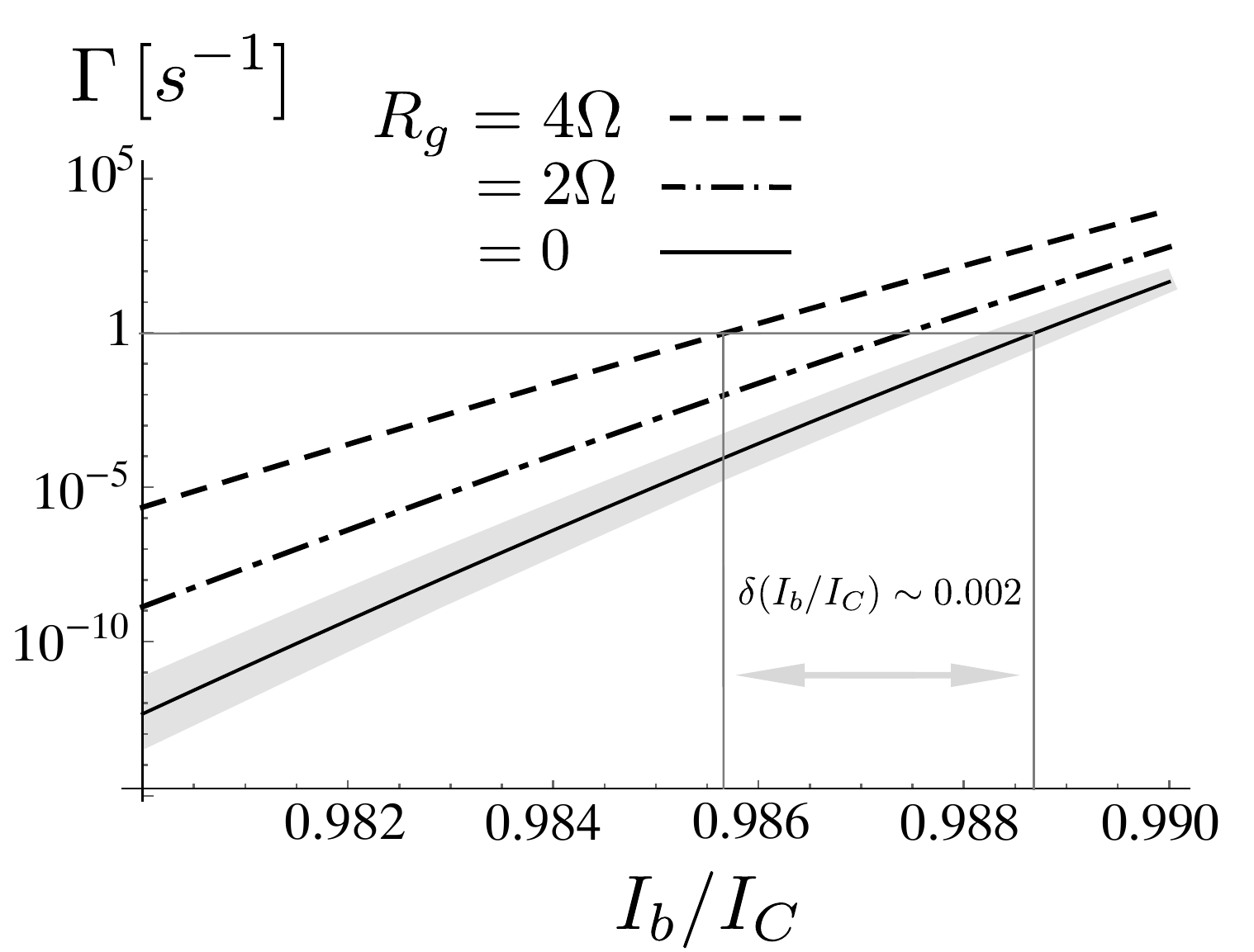}
	\caption{$\Gamma$ as obtained by using $K=K_0$ for the prefactor, for $C_J/C=0.02$, $R_S = 100\,\Omega$, and different values of $R_g$. 
	The other parameters are: $I_C=21\,\mu$A and $C_{tot}=6$ pF. 
	For 
	$R_g=0$, we also show the uncertainty induced by a $10\,\%$ experimental uncertainty of the capacitance $C_{tot}$ (gray area). 
	Increasing $R_g$ leads to a significant enhancement of the escape rate. For larger $I_b/I_C$ the increase due to charge dissipation becomes smaller, as can be seen from the three lines getting closer. For the observation of the effect, the ratio $I_b/I_C$ must be determined with an accuracy of the order of $\delta(I_b/I_C)<0.002$ for $R_g=4\,\Omega$. }
	\label{Fig.Rate}
\end{figure}

In order to estimate the escape rate $\Gamma$, 
we approximate the prefactor in Eq.~(\ref{Eq.decay2}) with $K \simeq K_0$
and use $K_0=a_q\omega_I/(2\pi)$, where $a_q \approx 52.1\sqrt{ V_{0}/(\hbar\omega_{I})}$ \cite{devoret_measurements_1985}. 
In a previous work (see Ref.~\cite{maile_effects_2020}), we found that the prefactor follows the qualitative behavior of enhancement/suppression for charge/phase dissipation, respectively. 
Hence, by assuming $ K$ to be independent of the dissipation,  we find a lower bound of the effect of dissipation for the case of pure charge and pure phase dissipation.
The escape rate $\Gamma$ obtained in this way is reported in Fig.~\ref{Fig.Rate}, as a function of $I_b/I_C$ and for different values of $R_g$, with the parameters $C_J/C=0.02$ and $R_S = 100\,\Omega$. 
%
%
%
%
%
%
%
%
For a realistic experimental setup, the rate should not be too small and also not too large, hence the bias current should be in a parameter range where $\Gamma \in [10^{-3}, 1] $ s$^{-1}$ yielding $I_b/I_C \in [0.985,0.988]$.
We find that for larger $I_b/I_C$ the three lines become closer to each other. This is reminiscent of the 
barrier dependence 
of the respective dissipative couplings discussed above.
The solid line in Fig.~\ref{Fig.Rate} shows the rate without charge dissipation and the surrounding gray area the experimental uncertainty of the capacitance $\delta C/C \sim 10\, \% $.
We find that the rate affected by charge dissipation exceeds 
this uncertainty interval.
However, the main obstacle for the observability of the rate enhancement introduced by the resistance $R_g$ is represented by the experimental uncertainty of $I_C$.
In fact, 
the rate of the circuit for a resistance $R_g=4\,\Omega$  
reaches the value $1$ s$^{-1}$ around $I_b/I_C \sim 0.986$ (dashed line).
This is the same value of the curve $R_g=0$ but with a bias curent 
$I_b/I_C \sim 0.988$ (solid line), namely when 
the real current bias may be shifted to larger values 
respect to the the nominal current in the experiment.
This implies that the ratio between $I_b$ and $I_C$ must be determined within an accuracy of $\delta(I_b/I_C) \sim 10^{-4}$.

With these premises, it is not surprising that the described effect has not been observed in the experiments done so far.
Although $R_g$ leads to a substantial enhancement, its effect might easily be absorbed into the experimental uncertainties without further notice.
More importantly, the here described effect of charge dissipation is only accessible if $R_g$ couples to the phase-coherent part of the junction.
Hence, such a resistor must be localized very close to the Josephson system and generate a distinct influence compared to other resistors and noise sources located further away from the junction on the experimental Josephson circuit. 
%
%
%
%
%
In other words, to observe the described effect by varying $I_b$, one needs to manufacture two junctions with very accurately determined parameters and add the resistor close to each of them. However, the effect is clearly observable if one engineers a tuneable resistance in order to study the dependence on $R_g$ directly (in situ).

For larger $R_g$ also the change in the functional dependence of the rate on $I_b/I_C$ becomes more substantial and might offer a possibility to measure the effects of $R_g$.

\setlength{\tabcolsep}{.12cm}
\begin{table}[b]
\begin{tabular}{| c |c | c | c | c | c || c |}
\hline
  $I_C$ & $C_J$ & $I_b$  &$C$ & $R_S$ & $R_g$ & $\mathcal{E}$ \\
 \hline
  $21\,\mu$A & $0.12\,$pF & $20.69\,\mu$A & $5.88\,$pF & $100\,\Omega$ & $4\,\Omega$ & $ 2.1\cdot 10^4 $ \\ 
 \hline
\end{tabular}
 \caption{Parameters used in Fig.~\ref{Fig.Rate} to explain the observability of the effect ($I_b/I_C=0.985$, $C_{tot}=6\,$pF, $C_J/C=0.02$). We choose a moderate $R_S$ to model unwanted dissipative effects on the phase variable. With respect to the non dissipative case the enhancement of the escape rate is $\mathcal{E}\sim 10^4$, specifically the rate changes from $\Gamma \sim 10^{-5}\,\text{s}^{-1}$ to $\Gamma \sim 10^{-1}\,\text{s}^{-1}$  .  } 
 \label{tab1}
\end{table}

%
%
%
\section{Conclusions} 
\label{Sec.5}
We theoretically studied a Josephson junction circuit in presence of two different kind of dissipative environments.
Specifically, we investigate a dissipative interaction that enhances the phase fluctuations - that we denoted charge dissipation -  in combination with the usual dissipative interaction that suppresses such fluctuations and that acts in favor of the localization of the phase.
These two kinds of dissipation are determined by the two resistances, $R_S$ and $R_g$, shown in the circuit of Fig.~\ref{Fig.1}.
We analyze their effect on the escape rate of the phase from a metastable state 
of a current-biased Josephson junction.
Considering realistic circuit parameters, we find that the escape rate can be enhanced by the dissipation affecting the charge when such dissipative interaction is dominant.

The experimental observability of the described effect strongly depends on the circuit parameters. 
For this reason, we estimate a possible parameter space in Sec.~\ref{Sec.4} for which the effect is measurable and suggest specific values in table \ref{tab1}.

Although we have analyzed the enhancement of the
escape rate due to the environmental assisted tunneling in a specific experimental 
system, such effects can be observed even in other platforms.
For example, the dissipative stabilization of quadrature squeezing has been recently experimentally reported \cite{dassonneville_dissipative_2021}.
There the enhanced uncertainty/fluctuations of the anti-squeezed quadrature could qualitatively yield the same effect described here, leading 
to a speedup in the tunneling process from a metastable well.

\acknowledgments
We are grateful to R. Dassonneville and J. Pekola 
for fruitful discussions.
We acknowledge financial support from the MWK-RiSC program.
This research was also partially supported by the German Excellence
Initiative by the Deutsche
Forschungsgemeinschaft (DFG) through the SFBs 767 and 1432, as well as the Project-ID 425217212.\\

\appendix


\section{Modelling quantum dissipation via transmission lines}
\label{Ap:dissipation}
\begin{figure}[h!]
	\centering
	\includegraphics[scale=0.75]{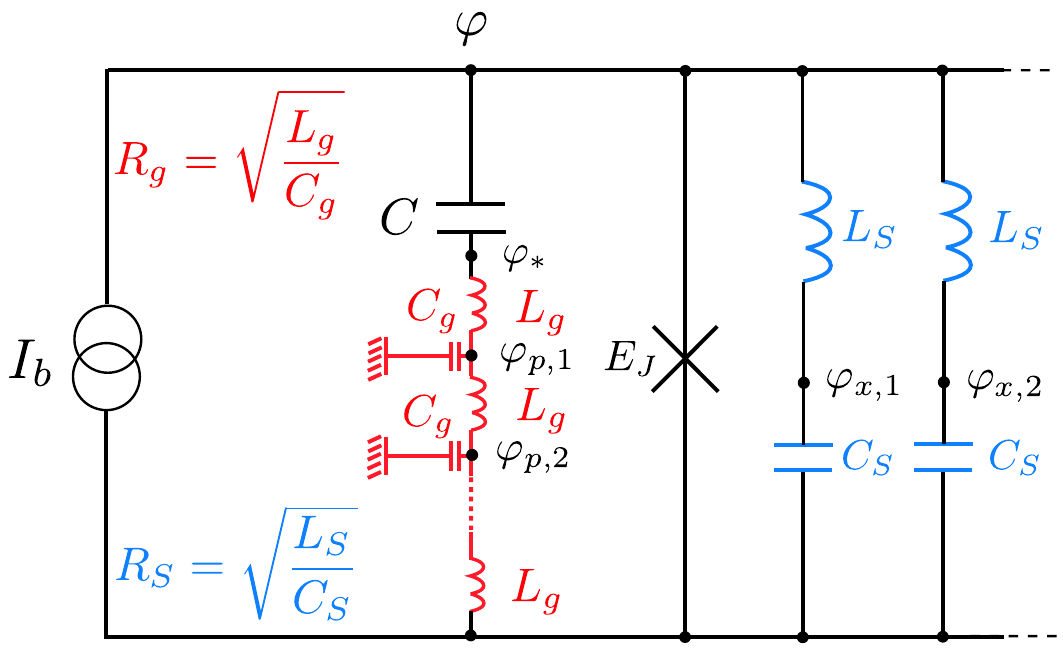}
	\caption{Transmission line circuit used to calculate the dissipative kernels for the Ohmic resistors in Fig.~\ref{Fig.1}.}
	\label{Fig.AppA}
\end{figure}

In this Appendix, we give some details of the theoretical origins of the Ohmic kernels, Eqs.~\eqref{Eq.phasekernel} and \eqref{Eq.chargekernel}.
We discuss the more fundamental circuit in Fig.~\ref{Fig.AppA}, where we neglect the capacitance of the junction $C_J$ for simplicity.

\subsection{Phase dissipation}
The resistor affecting the phase can be interpreted as an admittance shunting the Josephson junction. 
The dissipative kernel Eq.~\eqref{Eq.phasekernel}  can be found via coupling an infinite number of LC-resonators in parallel to the junction (blue circuits in Fig.~\ref{Fig.AppA}) \cite{Devoret2017}. 
Denoting the inductances of the resonators $ L_S $ and the capacitances $ C_S $ we find the Caldeira-Leggett type Euclidean Lagrangian 
\begin{align}
\mathcal{L}_B^{(\varphi)}=\frac{\Phi_0^2}{8\pi^2}\sum_{i=0}^{M-1} \left(C_S\dot{\varphi}_{x,i}^2+\frac{\left(\varphi_{x,i}-\varphi\right)^2}{L_S}\right),
\end{align}
where the bath degrees of freedom $ \varphi_{x,i} $ couple to the phase operator $ \varphi $. Integrating out the phases of the bath and letting $M \rightarrow \infty$, we insert phenomenologically the Ohmic spectral density describing the resistance $R_S$ and obtain the standard dissipative kernel in Eq.~(\ref{Eq.phasekernel})  \cite{Weiss2012,maile_quantum_2018}.

\subsection{Charge dissipation}
Since the dissipative kernel to the charge is by far less explored as compared to the 
coupling to the phase, we provide a more detailed explanation here (for further details see Refs.~\cite{maile_quantum_2018,Maile2021_thesis}).
To obtain the kernel Eq.~\eqref{Eq.chargekernel}, we couple a transmission line consisting of capacitances $C_g$ and inductances $L_g$ (see Fig.~\ref{Fig.AppA}).

The corresponding Euclidean
Lagrangian (system and charge environment) is $\mathcal{L}=\mathcal{L}_{S}+\mathcal{L}^{(Q)}_{B}$,
with 
\begin{equation}
\mathcal{L}_{S}=\frac{C\Phi_{0}^{2}}{8\pi^2}(\dot{\varphi} -\dot{\varphi}_{*})^{2}-E_{J}\cos\left(\varphi\right)- \frac{I_{b}\Phi_{0}}{2\pi}\varphi
\, .
\end{equation}
We emphasize that the time derivative of the phase $\dot{\varphi}$ is linearly coupled to the time derivative of the phase $\dot{\varphi}_*$ of the chain.
This term represent the charge energy associated to the capacitance $C$.
The Lagrangian  $\mathcal{L}^{(Q)}_{B}$ reads 
\begin{align}
\mathcal{L}^{(Q)}_{B}&=
\frac{\Phi_{0}^{2}}{8\pi^2L_{g}}(\varphi_{*}-\varphi_{p,1})^{2}\nonumber\\&+\sum_{n=1}^{M-1}\left[\frac{C_{g}\Phi_{0}^{2}}{8\pi^2}\dot{\varphi}_{p,n}^{2}+\frac{\Phi_{0}^{2}}{8\pi^2L_{g}}(\varphi_{p,n}\!-\!\varphi_{p,n+1})^{2}\right],
\end{align} 
where the dissipative environment $ \varphi_{p,n}  $ affects the phase $ \varphi_* $, which in turn is capacitively coupled to the charge $ \dot{\varphi} $.
To simplify the discussion, in contrast to the main text, we calculate the partition function and not density matrix elements in this Appendix. Within the path integral formalism this yields
\begin{align}
    Z\!=\!\oint\! D[\varphi]\!\oint \!D[\varphi_{*}]e^{-\frac{1}{\hbar}\int_{0}^{\beta}\mathcal{L}_{S}d\tau}\underline{\prod_{n}\!\oint\! D[\varphi_{p,n}]}e^{-\frac{1}{\hbar}\int_{0}^{\beta}\mathcal{L}^{(Q)}_{B}d\tau},\label{Eq.A4}
\end{align}
where the underline denotes that the product only acts on the integrals and not on the exponential.  To obtain the kernel for momentum dissipation, we have to perfom the last part of Eq.~(\ref{Eq.A4}) - which corresponds to \textit{integrating out} the bath degrees of freedom - and  also the integrals over the paths $\varphi_*(\tau)$.
In order to perform this integration, we diagonalize the semi-infinite transmission line via the eigenfunctions
\begin{align}
\varphi_{p,n}=\sqrt{\frac{2}{M}}\sum_{k=1}^{M-1}\sin\left(\frac{\pi k\;n}{M}\right)\phi_{k}
\end{align}
yielding 
\begin{align}
    \mathcal{L}^{(Q)}_B=&\frac{\Phi_{0}^{2}}{8\pi^2L_{g}}\!\left[\varphi_{*}^{2}-2\varphi_{*}\sum_{k=1}^{M-1}\epsilon_{k}\phi_{k}\right] \\ \nonumber &+\frac{\Phi_{0}^{2}C_{g}}{8\pi^2}\sum_{k=1}^{M-1}\left[\dot{\phi}_{k}^{2}+\Omega_{k}^{2}\phi_{k}^{2}\right],
\end{align}
where $\epsilon_k=\sqrt{2/M}\sin\left(\frac{\pi k n}{M}\right)$, $\Omega_{k}^{2}=4\omega_{g}^{2}\sin^{2}\left(\frac{\pi k}{2M}\right)$ and $\omega_g=\sqrt{1/L_{g}C_{g}}$.
Performing a Fourier transform from imaginary time to Matsubara frequencies $\omega_l=2\pi l/\beta$ (where $\beta = \hbar/(k_B T)$), we find for the corresponding action of the bath
\begin{align}
\int_{0}^{\beta}d\tau\mathcal{L}^{(Q)}_{B} = S_{*}+S_{B,0}+S_{B,1}^{(Re)}+S_{B,1}^{(Im)},
\end{align}
where 
\begin{align}
    S_{*}=\frac{\Phi_{0}^{2}}{4\pi^2L_{S}}\beta\left(\frac{\varphi_{*,0}^{2}}{2}+\sum_{l=1}^{\infty}|\varphi_{*,l}|^{2}\right),
\end{align}
is the part only containing $\varphi_*$,
\begin{align}
   S_{B,0} = & \beta\frac{\Phi_{0}^{2}C_{S}}{8\pi^2}\sum_{k=1}^{M-1}\Omega_{k}^{2}\left(\phi_{k,0}-\omega_{0}^{2}\epsilon_{k}\varphi_{*,0}\right)^{2}\nonumber\\&-\beta\frac{\Phi_{0}^{2}C_{S}}{8\pi^2}\sum_{k=1}^{M-1}\frac{\omega_{0}^{4}\epsilon_{k}^{2}}{\Omega_{k}^{2}}\varphi_{*,0}^{2}
\end{align}
and the parts $S^{(Re)}_{B,1}$ and $S^{(Im)}_{B,1}$ read
\begin{align} 
    S^{(j)}_{B,1} \!= & \frac{\beta\Phi_{0}^{2}C_{g}}{4\pi^2}\!\sum_{k=1}^{M-1}\sum_{l=1}^{\infty}\!\left(\omega_{l}^{2}\!+\!\Omega_{k}^{2}\right)\!\left[\phi_{k,l}^{(j)}\!-\!\frac{\omega_{0}\epsilon_{k}}{(\omega_{l}^{2}+\Omega_{k}^{2})}\varphi_{*,l}^{(j)}\right]^{2}\nonumber\\&-\frac{\beta\Phi_{0}^{2}C_{g}}{4\pi^2}\sum_{k=1}^{M-1}\sum_{l=1}^{\infty}\frac{\omega_{0}^{4}\epsilon_{k}^{2}}{(\omega_{l}^{2}+\Omega_{k}^{2})}(\varphi_{*,l}^{(j)})^{2}.
\end{align}
In this form, we can integrate out the phases $\phi_k$. The path integral measure transforms to Matsubara space via
\begin{align}
    \oint D[\phi_{k}]\rightarrow\int\frac{d\phi_{k,0}}{\sqrt{\frac{8\pi^3\hbar\beta}{\Phi_{0}^{2}C_{g}}}}\prod_{l=1}^{\infty}\int\frac{d\phi_{k,l}^{(Re)}d\phi_{k,l}^{(Im)}}{\frac{4\pi^3\hbar}{\beta\Phi_{0}^{2}C_{g}\omega_{l}^{2}}}.
\end{align}
After performing the path integral, we obtain the effective action for $\varphi_*$
\begin{align}
S_{B,eff} = \frac{\Phi_{0}^{2}\beta}{8\pi^2L_{g}}\frac{\varphi_{*,0}^{2}}{M}+\frac{\Phi_{0}^{2}\beta}{4\pi^2L_{g}}\sum_{l=1}^{\infty}|\varphi_{*,l}|^{2}\left(\frac{1}{M}+\omega_{l}Y_{l}\right),
\end{align}
where we defined the admittance
\begin{align}
    Y_{l}&=2\frac{\omega_{l}}{M}\sum_{k=1}^{M-1}\frac{\left(1-\frac{\Omega_{k}^{2}}{4\omega_{g}^{2}}\right)}{(\omega_{l}^{2}+\Omega_{k}^{2})}.\label{Eq.A12}
\end{align}
Substituting $x_{l}=\frac{\omega_{l}}{2\omega_{g}}$ and $\vartheta = \frac{\pi k}{2M}$, we can rewrite this and find in the limit $M\rightarrow \infty$ the integral
\begin{align}
    Y_{l}&=\frac{x_{l}}{\omega_{g}}\frac{2}{\pi}\int_{0}^{\frac{\pi}{2}}d\vartheta\frac{1-\sin^{2}\left(\vartheta\right)}{(x_{l}^{2}+\sin^{2}\left(\vartheta\right))}=\frac{1}{\omega_{g}}f_{c}(\omega_{l}),
\end{align}
where $f(\omega_l)$ is a high frequency cutoff function. 
Performing the limit of $M\rightarrow\infty$, the first two terms in Eq.~(\ref{Eq.A12}) vanish and we find that the effective action affecting $\varphi_{*}$ reads
\begin{align}
   S_{B,eff} &=  \frac{\Phi_{0}^{2}\beta}{4\pi^2L_{g}}\sum_{l=1}^{\infty}|\varphi_{*,l}|^{2}\omega_{l}Y_{l}\\
   & = \frac{\hbar}{2\pi}\frac{R_{q}}{R_{g}}\beta\sum_{l=1}^{\infty}|\varphi_{*,l}|^{2}\omega_{l}f_{c}(\omega_{l}),
\end{align}
where we used that $\sqrt{L_g/C_g}$ is the characteristic impedance of the semi-infinite transition line acting as an Ohmic resistor and therefore introduced $R_g=\sqrt{L_g/C_g}$. Further,
$R_{q}=h/(4e^2)$ is the resistance quantum. We will set $f(\omega_l)=1$ in the following, because its contribution is irrelevant for the problem at hand. The remaining path integral to solve reads
\begin{align}
    Z\!=\!\oint\! D[\varphi]\!\oint \!D[\varphi_{*}]e^{-\frac{1}{\hbar}\int_{0}^{\beta}\mathcal{L}_{S}d\tau+S_{B,eff}},
\end{align}
where the first part in the exponential contains the Lagrangian of the charging part $\mathcal{L}_{S}^{(1)}=C\Phi_{0}^{2}(\dot{\varphi}-\dot{\varphi}_{*})^{2}/(8\pi^2)$. We again perform a transformation to Matsubara frequencies for $\varphi$ and $\varphi_*$ an integrate out the latter one. With this we obtain the remaining action
\begin{align}
S_{R} & =\frac{C\Phi_{0}^{2}}{4\pi^2}\beta\sum_{l=1}^{\infty}\omega_{l}^{2}|\varphi_{l}|^{2}-\frac{C\Phi_{0}^{2}}{4\pi^2}\beta\sum_{l=1}^{\infty}\frac{\tau_{p}\omega_{l}^{3}}{\left[1+\tau_{p}\omega_{l}\right]}|\varphi_{l}|^{2}, \label{App.effectact}
\end{align}
where we introduced $\tau_{p}=CR_g$. The action in Eq.~(\ref{App.effectact}) is the effective action of the phase variable of the system affected by the resistor $R_g$ and therefore the Fourier transformed version of Eq.~(\ref{eq:Saction}) in Matsubara frequencies for pure charge dissipation. Hence, we find for the dissipative kernel in Eq.~(\ref{Eq.chargekernel})
\begin{equation}
F^{(Q)}(\tau)=-\frac{C\Phi_{0}^{2}}{8\pi^2\beta}\sum_{l=-\infty}^{\infty}\frac{\tau_{p}|\omega_{l}|}{1+\tau_{p}|\omega_{l}|}e^{i\omega_{l}\tau}.
\end{equation}

\section{Discussion between the numerical and analytical results}
In Sec.~\ref{Sec.2}, we discussed different techniques to calculate the bounce path minimizing the action. 

First, there is the approximation of Caldeira and Leggett. They treat the weak-coupling limit and insert the non-dissipative solution for the bounce path into the action to obtain approximate results. In case of pure phase dissipation, this appears to be an excellent approximation as shown in Fig.~\ref{Fig.Fig3}c. Therefore, it is not surprising that also the numerical results and the variational results coincide in the parameter space considered in this work.  
However, as we discuss in the main text, using the non-dissipative bounce underestimates the effect of charge dissipation as we saw in Figs.~\ref{Fig.Fig3} and \ref{Fig.Fig6}.  

%
%
%
%
%
%
%
\begin{figure}[t]
	\centering
	\includegraphics[scale=0.25]{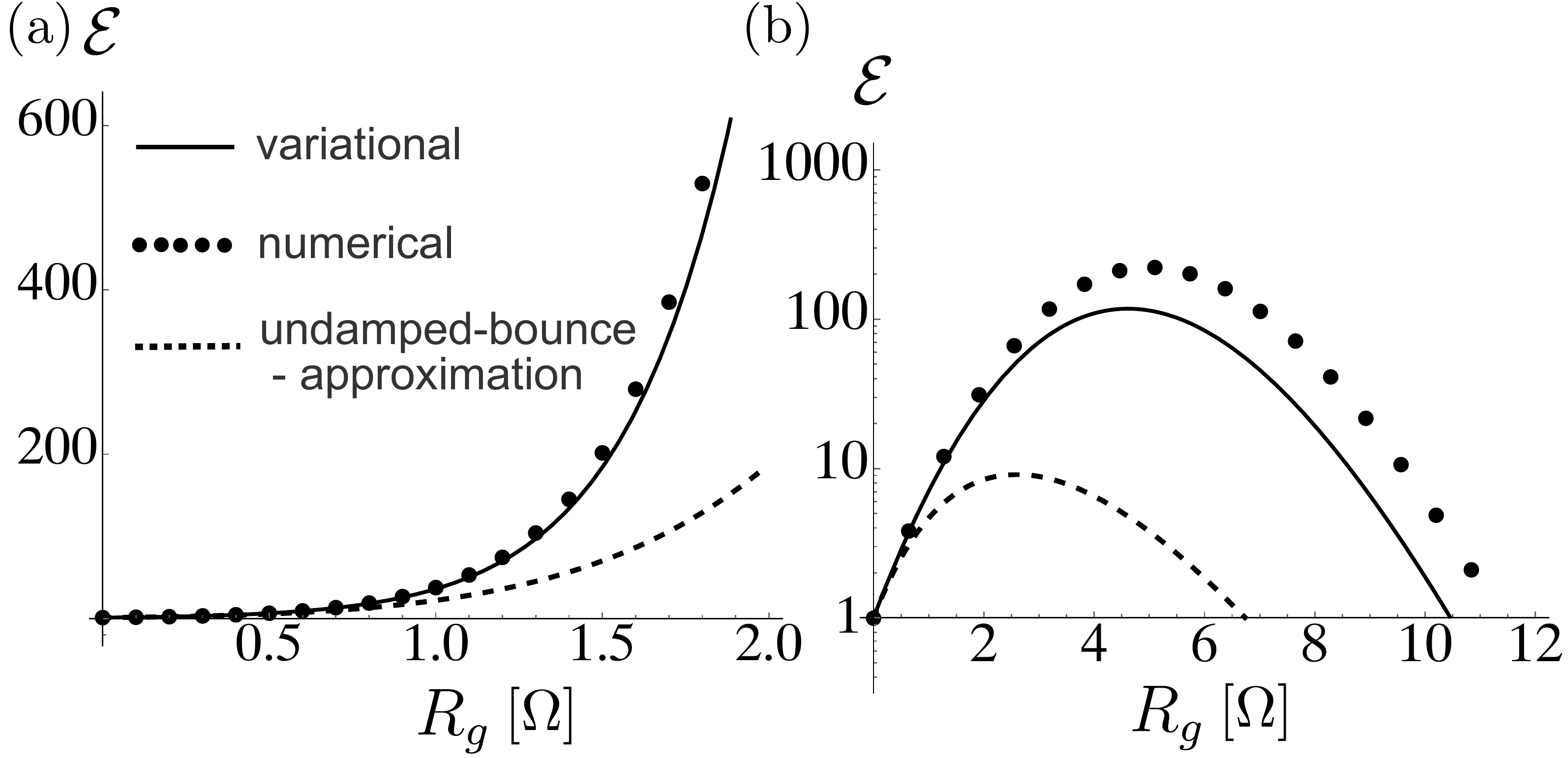}
	\caption{Comparison of the three different techniques to calculate the action
	minimized by the bounce path. We show result for the exact numerical (dots) and the variational (solid line) results, together with the solution using the non-dissipative bounce approximation (dashed line). (a) Pure charge dissipation.  (b) Both dissipative couplings the variational treatment.}
	\label{Fig.App2}
\end{figure}
%
%
%
%
%
%
Here, we show a comparison between the variational and the numerical treatment. In Fig.\ref{Fig.App2}a, we see that the variational treatment yields a very good approximation to the exact numerical result in case of pure charge dissipation, in the considered parameter range. As expected, the quality of the approximation depends on the coupling strength ($\propto R_g$ in our case). 
In presence of both dissipative couplings, we use larger values of $R_g$ in Fig.~\ref{Fig.boxplots}b.
We show the differences between the three techniques in Fig.~\ref{Fig.App2}b.
For larger $R_g$, the variational treatment becomes less accurate, although the qualitative behavior and the order of magnitude are still in good agreement. 
Using the non-dissipative bounce strongly underestimates the effect of charge dissipation.
The only plot in the main text that yields a noticeable difference in the results is Fig. \ref{Fig.boxplots}b, as $R_g$ is of the order of $10~\Omega$. We show in Fig.~\ref{Fig.App2}b that the variational treatment yields result of the same order of magnitude (in contrast to the Caldeira-Leggett approxmation). Further, we see that the variational treatment underestimates the exact result and can be seen as a lower bound of the effect of $R_g$. The parameters used in (b) coincide with the blue solid curve in Fig.~\ref{Fig.boxplots}b.

\bibliographystyle{mybibstyle}
 
\bibliography{bibliography}

\end{document}